\documentclass[aps,pra,twocolumn,amsmath,amssymb,nofootinbib,showpacs,superscriptaddress]{revtex4-1}
\usepackage[english]{babel}
\usepackage{latexsym}
\usepackage{graphics}
\usepackage{graphicx}
\usepackage{epsfig}
\usepackage{color}
\usepackage{bm}
\usepackage{amsmath}
\usepackage{amssymb}
\usepackage{amsthm}
\usepackage{dcolumn}
\usepackage{bm}
\usepackage{float}
\usepackage{hyperref}
\usepackage{color}
\usepackage{epstopdf}
\usepackage[svgnames]{xcolor}
\usepackage{enumerate}
\usepackage[svgnames]{xcolor}
\hypersetup{hidelinks,colorlinks=true,allcolors=DarkBlue}

\newcommand{\kpr}{k_{\rm sec}}
\newcommand{\kpb}{k_{\rm publ}}
\newcommand{\sgn}{{\rm sgn}}

\begin{document}

\preprint{APS/123-QED}

\title{Quantum-secured blockchain}
\author{E.O. Kiktenko} 
\affiliation{Russian Quantum Center, Skolkovo, Moscow 143025, Russia}
\affiliation{Steklov Mathematical Institute of Russian Academy of Sciences, Moscow 119991, Russia}

\author{N.O. Pozhar} 
\affiliation{Russian Quantum Center, Skolkovo, Moscow 143025, Russia}

\author{M.N. Anufriev} 
\affiliation{Russian Quantum Center, Skolkovo, Moscow 143025, Russia}

\author{A.S. Trushechkin}
\affiliation{Russian Quantum Center, Skolkovo, Moscow 143025, Russia}
\affiliation{Steklov Mathematical Institute of Russian Academy of Sciences, Moscow 119991, Russia}

\author{R.R. Yunusov}
\affiliation{Russian Quantum Center, Skolkovo, Moscow 143025, Russia}

\author{Y.V. Kurochkin}
\affiliation{Russian Quantum Center, Skolkovo, Moscow 143025, Russia}

\author{A.I. Lvovsky}\email{LVOV@ucalgary.ca}
\affiliation{Russian Quantum Center, Skolkovo, Moscow 143025, Russia}
\affiliation{Institute for Quantum Science and Technology, University of Calgary, Calgary AB T2N 1N4, Canada}

\author{A.K. Fedorov}\email{akf@rqc.ru}
\affiliation{Russian Quantum Center, Skolkovo, Moscow 143025, Russia}

\date{\today}
\begin{abstract}
Blockchain is a distributed database which is cryptographically protected against malicious modifications. 
While promising for a wide range of applications, current blockchain platforms rely on digital signatures, which are vulnerable to attacks by means of quantum computers. 
The same, albeit to a lesser extent, applies to cryptographic hash functions that are used in preparing new blocks, 
so parties with access to quantum computation would have unfair advantage in procuring mining rewards. 
Here we propose a possible solution to the quantum-era blockchain challenge and report an experimental realization of a quantum-safe blockchain platform 
that utilizes quantum key distribution across an urban fiber network for information-theoretically secure authentication. 
These results address important questions about realizability and scalability of quantum-safe blockchains for commercial and governmental applications.

\end{abstract}
\maketitle

\section*{Introduction}

The blockchain is a distributed ledger platform with high Byzantine fault tolerance, which enables achieving consensus in a large decentralized network of parties who do not trust each other. 
A paramount feature of blockchains is the accountability and transparency of transactions, 
which makes it attractive for a variety of applications ranging from smart contracts and finance to manufacturing and healthcare~\cite{Franco}. 
One of the most prominent applications of blockchains is cryptocurrencies, such as Bitcoin~\cite{Extance2015}. 
It is predicted that ten percent of global GDP will be stored on blockchains or blockchain-related technology by 2025 \cite{Forbes}.

In a modern blockchain network, any member can introduce a record (\emph{transaction}) to the ledger. 
Every transaction must be signed by its initiator's digital signature; this rule enables, for example, exchange of digital assets between parties.
The transactions are stored on each member's computer (\emph{node}) as a sequence of groups known as \emph{blocks}. 
All transactions that have been introduced over a period of time are compiled in a block that is linked to the previous one~\cite{Swan2015}. 
This linking is implemented by cryptographic hash functions: each block contains a hash value of its content, and the content also includes the hash of the previous block (Fig.~1). 
Any modification of a block inside the chain yields a change of its hash, which would in turn require modification of all subsequent blocks. 
This structure protects the data inside a blockchain from tampering and revision~\cite{Witte2016}. 

\begin{figure}[h]
	\includegraphics[width=0.9\columnwidth]{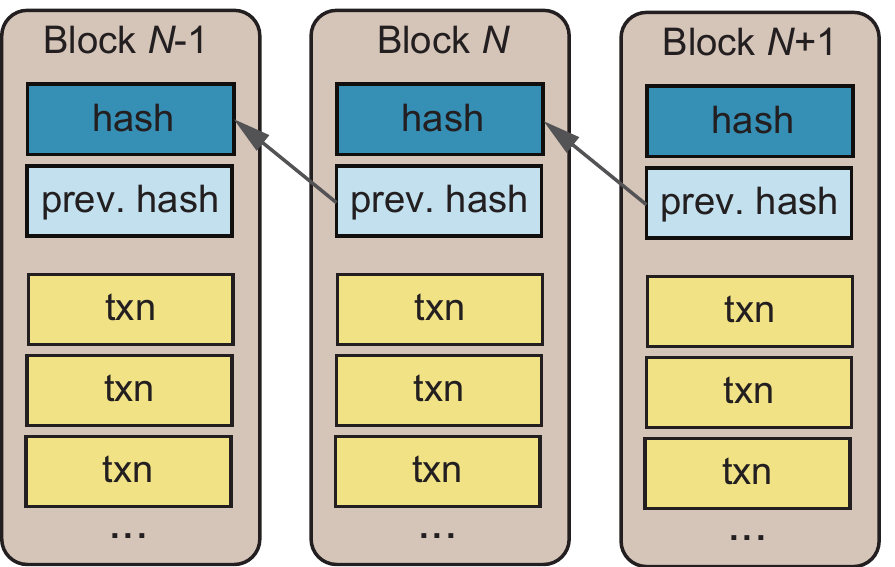}
	\vskip -2mm
	\caption{Organization of data in blockchains (``txn" stands for ``transaction").}
	\label{fig:QKP}
\end{figure}

While each node is allowed, in principle, to introduce a block to the network, each blockchain network has a set of rules that organize and moderate the block formation process. 
In Bitcoin~\cite{Extance2015}, for example, a member introducing a new block must solve an NP-hard problem: 
introduce a set of numbers to the block's header such that the hash of that header must not exceed a certain value (this paradigm is known as \emph{proof-of-work}). 
In this way, the blocks are guaranteed not to emerge too frequently, so every node has an opportunity to verify the validity of the block and the transactions therein before a new block arrives. 
This ensures the identity of the database stored by all network nodes. 
Whenever a new block is accepted by the community, its ``miner" is rewarded in bitcoins for the computational power they spend.  

A more detailed summary of the blockchain concept is presented in Appendix A. 

We see that blockchain relies on two one-way computational technologies: cryptographic hash functions and digital signatures. 
Most blockchain platforms rely on the elliptic curve public-key cryptography (ECDSA) or the large integer factorization problem (RSA) to generate a digital signature~\cite{Witte2016}. 
The security of these algorithms is based on the assumption of computational complexity of certain mathematical problems~\cite{Schneier1996}.

A universal quantum computer would enable efficient solving of these problems, thereby making corresponding digital signature algorithms, including those used in blockchains, insecure. 
In particular, Shor's quantum algorithm solves factorisation of large integers and discrete logarithms in polynomial time~\cite{Shor1997}.
Another security issue is associated with Grover's search algorithm~\cite{Grover1996}, which allows a quadratic speedup in calculating the inverse hash function. 
In particular, this will enable a so-called 51-percent attack, in which a syndicate of malicious parties controlling a majority of the network's computing power would monopolize the mining of new blocks. 
Such an attack would allow the perpetrators to sabotage other parties' transactions or prevent their own spending transactions from being recorded in the blockchain.
Other attacks with quantum computing on blockchain technology as well as possible roles of quantum algorithms in the mining process are considered in more detail in recent publications~\cite{Tessler2017,Aggarwal2017,Kalinin2018,Sapaev2018}.

The security of blockchains can be enhanced by using post-quantum digital signature schemes \cite{Lamport,Merkle,Bernstein2009} for signing transactions. 
Such schemes are considered to be robust against attacks with quantum computers~\cite{Bernstein2009}. However, this robustness relies on unproven assumptions. 
Furthermore, post-quantum digital signatures are computationally intensive and are not helpful against attacks that utilize the quantum computer to dominate the network's mining hashrate.

In addition to the blockchains based on mining principles there are other approaches to distributed ledgers maintenance, e.g. Byzantine fault tolerance (BFT) replication~\cite{Lamport1982} and practical BFT replication~\cite{Castro2002}.
To our knowledge, all the proposed approaches either require use of digital signatures, and hence are vulnerable to quantum computer attacks, or pairwise authenticated channels at least.
We note that the pairwise authentic channel ensures that each message was not tampered while passing, but does not solve the transferability issue."

The way to guarantee authentication in the quantum era is to use quantum key distribution (QKD), 
which guarantees information-theoretic (unconditional) security based on the laws of quantum physics~\cite{Gisin2002,Scarani2009,Lo2016,Gyongyosi2018}. 
QKD is able to generate a secret key between two parties connected by a quantum channel (for transmitting quantum states) 
and a public classical channel (for post-processing procedures).
The technology enabling QKD networks have been demonstrated 
in many experiments~\cite{Laenger2009,Yeh2005,Peev2009,Stucki2011,Pan2009,Pan2010,Han2010,Zeilinger2011,Shields2013,Zhang2016,Pozhar2017} 
and is now publicly available through multiple commercial suppliers.

In the present work, we describe a blockchain platform that combines (i) the original BFT state-machine replication without use of digital signatures~\cite{Lamport1982} (hereafter referred to as the ``broadcast protocol''), (ii) QKD for providing authentication, and implement an experiment demonstrating its capability in an urban QKD network~\cite{Pozhar2017}.
We believe this scheme to be robust against not only the presently known capabilities of the quantum computer, 
but also those that may potentially be discovered in the future to make post-quantum cryptography schemes vulnerable.

The utility of QKD for blockchains may appear counterintuitive, as QKD networks rely on trust among nodes, whereas the earmark of many blockchains is the absence of such trust.
More specifically, one may argue that QKD cannot be used for authentication because it itself requires an authenticated classical channel for operation. 
However, each QKD communication session generates a large amount of shared secret data, part of which can be used for authentication in subsequent sessions. 
Therefore a small amount of ``seed" secret key that the parties share before their first QKD session ensures their secure authentication for all future communication \cite{Tysowski2017}.
In this way, QKD can be used in lieu of classical digital signatures.

\section*{Quantum-secured blockchain}

Here we consider a blockchain protocol within a two-layer network with $n$ nodes. 
The first layer is a QKD network with pairwise communication channels that permit establishing information-theoretically (unconditionally) secure private key for each pair of nodes. 
The second (classical) layer is used for transmitting messages with authentication tags based on information-theoretically secure Toeplitz hashing (see Appendix B) 
that are created using the private keys procured in the first layer.

For concreteness, we consider a blockchain maintaining a digital currency.
The operation of the blockchain is based on two procedures: (i) creation of transactions and (ii) construction of blocks that aggregate new transactions. 
New transactions are created by those nodes who wish to transfer their funds to another node. 
Each individual new transaction record is constructed akin to those in Bitcoin, 
i.e.~contains the information about the sender, receiver, time of creation, amount to be transferred, 
and a list of reference transactions that justifies that the sender has enough funds for the operation (see Appendix A).
This record is then sent via authenticated channels to all other $n-1$ nodes, thereby entering the pool of unconfirmed transactions. 
Each node checks these entries with respect to their local copy of the database and each other, 
in order to verify that each transaction has sufficient funds, and forms an opinion regarding the transaction's admissibility. 
At this stage, the community does not attempt to exclude double-spending events (a dishonest party sending different versions of a particular transaction to different nodes of the network).

Subsequently, the unconfirmed transactions are aggregated into a block. 
We abolish the classical blockchain practice of having the blocks proposed by individual ``miners", because it is vulnerable to quantum computer attacks in at least two ways. 
First, transactions are not rigged with digital signatures. 
This means that a miner has complete freedom to fabricate arbitrary, apparently valid, transactions and include them in the block. 
Second, a node equipped with a quantum computer is able to mine new blocks dramatically faster than any non-quantum node. 
This opens a possibility for attacks such as the 51-percent attack described above. 

Instead, we propose to create blocks in a decentralized fashion. 
To this end, we employ the broadcast protocol proposed in the classic paper by Shostak, Lamport and Pease~\cite{Lamport1982} (see Appendix C). 
This information-theoretically secure protocol allows achieving a Byzantine agreement 
in any network with pairwise authenticated communication provided that the number of dishonest parties is less than $n/3$ (which we assume to be the case). 
At a certain moment in time (e.g.~every ten minutes), the network applies the protocol to each unconfirmed transaction, 
arriving at a consensus regarding the correct version of that transaction (thereby eliminating double-spending) and whether the transaction is admissible. 
Each node then forms a block out of all admissible transactions, sorted according to their time stamps. 
The block is added to the database.
In this way, the same block will be formed by all honest parties, thereby eliminating the possibility of a ``fork'' -- the situation in which several different versions of a block are created simultaneously by different miners.

Because the broadcast protocol is relatively forgiving to the presence of dishonest or faulty nodes, 
our blockchain setup has significant tolerance to some of the nodes or communication channels not operating properly during its implementation. 
We also emphasize that, while the broadcast protocol is relatively data intensive, the data need not be transmitted through quantum channels. 
Quantum channels are only required to generate private keys.

While the proposed protocol seems to be efficient against quantum attacks on the distribution of transactions and formation of blocks, the database is still somewhat vulnerable while it is stored. 
A possible attack scenario is as follows: a malicious party equipped with a quantum computer works off-line to forge the database. 
It changes one of the past transaction records to its benefit and performs a Grover search for a variant of other transactions within the same block such that its hash remains the same, 
to make the forged version appear legitimate. 
Once the search is successful, it hacks into all or some of the network nodes and substitutes the legitimate database by its forged version. 
However, the potential of this attack to cause significant damage appears low, because the attacker would need to simultaneously hack at least one-third of the nodes to alter the consensus. 
Furthermore, because the Grover algorithm offers only a quadratic speed-up with respect to classical search algorithms, 
this scenario can be prevented by increasing the convention on the length of the block hash to about a square of its safe non-quantum value. 

We experimentally study the proposed blockchain protocol on the basis of a four-node, six-link network [Fig. 2(a)] with information-theoretically secure authentication. 
We use an urban fiber QKD network recently developed by our team (see Appendix D) to procure authentication keys for two of the links connecting three nodes; 
the key generation in the remaining four links is classical.
We sum up main parameters of the implemented blockchain for four nodes network in Table~\ref{tab:1}.
	
\begin{table}[H]
	\begin{tabular}{| p{0.78\linewidth} | p{0.18\linewidth} |} 
		\hline
		Number of nodes in the network & $n=4$\\ \hline
		Upper bound on the number of faulty nodes  & $m=1$\\ \hline
		Number of rounds in the broadcast protocol & $2$ \\ \hline
		Duration of broadcast protocol & $<10$ sec \\ \hline
		Time between block generation events & 5 min \\ \hline
		Authentication hash length & 40 bit \\ \hline
		Quantum key consumption in the initial broadcast of a transaction & 40 bit \\ \hline
		Quantum key consumption in the broadcast protocol & 80 bit \\ \hline
		Average quantum key consumption required for a transaction rate of 10 per minute &  $<7$ bit/s \\
		\hline
	\end{tabular}
	\caption{Main parameters of the implemented quantum-secured blockchain. 
	}
	\label{tab:1}
\end{table}

We test the operation of the blockchain and implement the construction of a simple transaction block under the following settings [Fig.~2(a)]. 
Nodes A, B and C perform legitimate transactions, whereas node D tries to process three different transactions, i.e.~realize a double-spending attack. 
The pool of unconfirmed transactions at each node thus consist of three legitimate and one inconsistent transactions. 
The broadcast protocol is then launched on the basis of these transaction pools. 
This protocol eliminates node D's double-spending transaction after the second communication round and permits the formation of a block containing legitimate transactions only [Fig.~2(c)].

\begin{figure}[h]
	\includegraphics[width=\columnwidth]{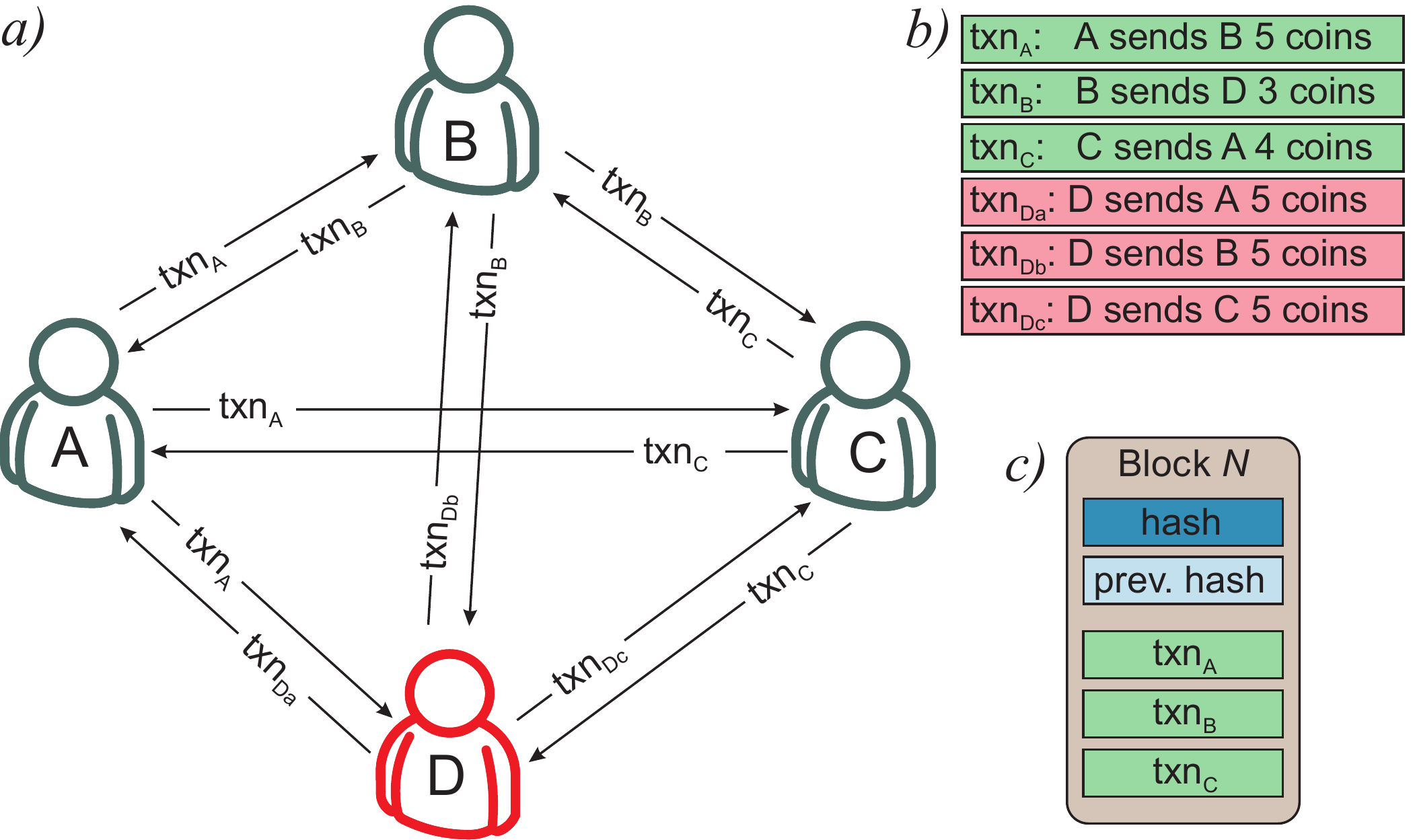}
	\caption{Creation of a block in a quantum-secure blockchain. 
	a) Each node who wishes to implement a transaction sends identical copies of that transaction to all other nodes. 
	Nodes A, B and C, whose transactions are denoted as txn$_{\rm A}$, txn$_{\rm B}$ and txn$_{\rm C}$, respectively, follow the protocol. 
	Node D is cheating, attempting to send non-identical versions txn$_{\rm Da}$, txn$_{\rm Db}$ and txn$_{\rm Dc}$ of the same transaction to different parties. 
	b) Transaction contents. 
	c) The nodes implement the broadcast protocol to reconcile the unconfirmed transactions and form the block. 
	They discover that the transaction initiated by node D is illegitimate and exclude it. }
\end{figure}

\section*{Conclusion and outlook}

In summary, we have developed a blockchain protocol with information-theoretically secure authentication based on a network in which each pair of nodes is connected by a QKD link. 
We have experimentally tested our protocol by means of  a three-party urban fibre network QKD in Moscow. 

In addition to using QKD for authentication, we have redefined the protocol of adding new blocks an a way that is dramatically different from modern cryptocurrencies. 
Rather than concentrating the development of new blocks in the hands of individual miners, 
we employ the information-theoretically secure broadcast protocol where all the nodes reach an agreement about a new block on equal terms. 

A crucial advantage of our blockchain protocol is its ability to maintain transparency and integrity of transactions against attacks with quantum algorithms.
Our results therefore open up possibilities for realizing scalable quantum-safe blockchain platforms. 
If realized, such a blockchain platform can limit economic and social risks from imminent breakthroughs in quantum computation technology. 

Typical key generation rates of currently available QKD technologies are sufficient for operating a large-scale blockchain platforms based on our protocol. 
Moreover, remarkable progress in theory and practice of quantum communications, including recent experiments on ground-to-satellite QKD and quantum repeaters, 
could open the door to developing a public worldwide QKD network (``the quantum Internet"~\cite{kimble}) and extending quantum-safe blockchain platforms to the global scale.  

The development of the ``quantum Internet" will allow our protocol to preserve anonymity of each network member. 
A member will be able to access the global QKD network from any station, authenticate themselves to other parties using their private seed keys (see Methods) and enact a desired transaction.

Our protocol is likely not the only possible quantum-safe blockchain platform. 
In this context, important horizons are opened by technologies that permit direct transmission of quantum states over multipartite communication networks combined with light quantum information processing. 
This includes, for example, 
protocols for quantum multiparty consensus~\cite{Gisin2001,Smania2016}, 
other approaches for QKD~\cite{Grangier2003,Gyongyosi2014,Gyongyosi,GyongyosiImre2018}, and quantum digital signatures~\cite{Gottesman2001}, 
which have been successfully studied in experiments, including metropolitan networks~\cite{Pan2017}. 
An additional important research avenue is more efficient, quantum-technology based consensus algorithms \cite{Gottesman2002} and general study of quantum channels~\cite{Holevo,GyongyosiImre2018}. 
Most importantly, we hope that our work will raise awareness and interest of the quantum information community to the problem of security of distributed ledgers in the era of quantum technology.

\section*{Acknowledgments} 
We thank D. Gottesman for making us aware of the broadcast protocol. 
This work is supported by the Russian Science Foundation under grant 17-71-20146.
AL is supported by NSERC and is a CIFAR Fellow.

\section*{Appendix A. Blockchain workflow}

Here we sum up the main definitions and concepts of conventional blockchains.
\begin{enumerate}[1.]
	\item The \emph{blockchain} is a distributed database in which the records are organized in a form of consecutive blocks.
	The term ``distributed" means that copies of the database are stored by all the nodes that are interested in maintaining it, and that there is no single control center in charge of the network.
	
	\item \emph{Distributed consensus} is a set of rules governing the blockchain construction and operation accepted by the nodes maintaining this blockchain.
	
	\item A \emph{transaction} is an elementary record in a blockchain.
	In order to create a transaction, one (i) forms a corresponding record, 
	(ii) signs it using a digital signature, 
	and (iii) sends the record to all the nodes maintaining the blockchain.
	For example, if we use a blockchain for maintaining a cryptocurrency, then the transaction corresponds to a transfer of some amount of money from one party to another.
	
	\item A \emph{block} contains a number of transactions created over a certain period of time. 
	Newly created transactions enter a so-called pool of unconfirmed transactions. 
	Because such transactions are created at a faster rate than the typical network latency time, it is difficult for the community to agree on their time sequence and validity. 
	This motivates the solution to aggregate new transactions into large blocks that are introduced at regular time intervals that are much longer than the network latency.
	
	In order to create a block with new transactions, a node needs to 
	(i) check the validity of new transactions and discard invalid ones, 
	(ii) combine the new transactions and the hash value of the last block in the existing blockchain, 
	(iii) fulfill the additional moderation requirements imposed on new proposed block by the network rules (an example is the proof of work rule in Bitcoin), 
	and (iv) broadcast the new block to all other nodes.
	Each node then verifies the block's validity and adds it to the local copy of the blockchain.
	
	\item A situation named \emph{fork} is possible when non-identical blocks are generated and broadcast by different miners at about the same time. In this situation, the community becomes temporarily divided as different miners will use these different blocks to generate their subsequent blocks. To reunite the community, the longest-chain rule applies: as the both branches continue to grow, one of them will become longer than others. At this point, the community chooses this particular branch as the ``correct'' one. As a consequence of this rule, the reliability of any block grows with its depth relative to the last block in the chain.
	
	\item The \emph{cryptographic hash function} $H(\cdot)$ is a one-way map from arbitrary length strings to fixed-length strings (let say, 256 or 512 bit). 
	The term ``cryptographic" means that it act is pseudo-random way, \emph{i.e.} any modification of the argument string $x$ (even in a single bit) yields a major and unpredictable change of $H(x)$.

	Moreover, it is universally believed that there is no classical algorithm, except brute-force, to invert the hash function (solve an equation $H(x)=h$ for a given hash $h$), and find its collisions (i.e. find the string $x\neq y$ for a given $y$ such that $H(x)=H(y)$, or even find two arbitrary distinct strings $x$ and $y$ such that $H(x)=H(y)$). 
	Quantum algorithms, in particular, Grover's algorithm~\cite{Grover1996}, allow a quadratic speedup in solving such problems only.

	\item A \emph{digital signature} is an algorithm that allows one to verify that a certain message (in our case, a transaction) has been created by a particular person.
	The basic idea is that the author generates a pair of keys: a secret key $\kpr$, which must be kept out of reach for all others, and a public key $\kpb$, which can be known to anyone. 
	There is a fixed-length output function $\sgn(x,k)$ taking an arbitrary message $x$ and a secret key $k$, such that the triplet 
	\begin{equation}
	\{m,\sgn(m,\kpr),\kpb\}
	\end{equation}
	verifies the fact the author, identified with the public key $\kpb$, indeed possesses the corresponding secret key $\kpr$ and signed the message $m$.
	On the other hand, the above triplet does not allow one to determine $\kpr$ using a reasonable amount of classical computational resources.
\end{enumerate}

\section*{Appendix B. Information-theoretically secure authentication}

Two parties, Alice and Bob, can authenticate messages sent to each other if they share a secret  private key $K_{\rm aut}$ that is not known to anyone else. 
The private key of necessary length can be generated via QKD provided that the parties have a small amount of ``seed" key to authenticate themselves to each other in the beginning of the session. 
Once the private key is established, the authentication procedure is as follows: Alice sends to Bob a message with a hash tag generated using that key. 
After receiving the message, Bob also computes its hash tag. If the hash tags coincide, Bob can be certain that the message has arrived from Alice. 

In our protocol, we use Toeplitz hashing due to its computational simplicity \cite{Krawczyk1994,Krawczyk1995}. 
Let the lengths of all messages and their hash tags be $l_M$ and $l_h$ respectively. 
The hash tag of the $i$th message $M_i$ is calculated according to
\begin{equation}\label{eq:aut}
	h(M_i)=T_SM_i\oplus r_i,
\end{equation}
where $T_S$ is a $l_h\times l_M$ Toeplitz matrix generated by a string $S$ of length $l_h+l_M-1$, $r_i$ is a bit string of length $l_h$, and $\oplus$ is the bitwise xor. 
Both $S$ and $r_i$ are private and taken from the common private key $K_{\rm aut}$. 
Then the probability that an eavesdropper will correctly guess the hash tag of a modified message is not more than $2^{-l_h}$. 

If a series of messages is transmitted, the string $S$ can be reused without compromising security, while the string  $r_i$ must be generated anew every time. 
In this way, the private key is consumed at a rate of $l_h$ bits per message. 
In our experiment, $l_h=40$ and $l_M=2^{22}$.

\section*{Appendix C. Broadcast protocol and block construction}

Here we briefly discribe the protocol for reaching Byzantine agreement in the presence of faulty nodes~\cite{Lamport1982}.
We consider $n$ nodes connected by pairwise authenticated channels.
Let each $i$th node possess a certain private value $V_i$.

The goal of the protocol is to make all nodes aware of all $V_i$'s  with a complication that  there are at most $m$ ``dishonest'' (or faulty) nodes.
This can be rephrased as obtaining an $n$-dimensional interactive consistency vector $\vec V^{\rm cons}$ with the following properties:
(i) all the honest nodes obtain the same vector $\vec V^{\rm cons}$, and (ii) the $i$th component of $\vec V^{\rm cons}$ equals $V_i$ for all honest nodes.

The interactive consistency vector is determined through a series of communication rounds that proceed as follows.
\begin{itemize}
	\item In the first  round, the nodes transmit their values of $V_i$ to each other.
	\item In subsequent rounds, the nodes communicate all the information they received in the previous round from other nodes 
	(messages are of the form such as ``node $i_2$ told node $i_1$ that node $i_3$ told node $i_2$ \ldots that node $i_r$ told node $i_{r-1}$ that its private value is $U$''). 
\end{itemize}
In Ref.~\cite{Lamport1982}, Lamport, Shostak and Pease proved that the interactive consistency vector can be obtained with no more than $m+1$ rounds for $m<n/3$. 

In our setup, the private value $V_i$ is the pool of transactions received by the $i$th node (together with its own transactions), 
as well as the set of bits indicating the node's opinion of each transaction's admissibility. 
After obtaining the interactive consistency vector $\vec V^{\rm cons}$, the honest nodes are able to create a block containing the complete set of admissible transactions from the pool.

A shortcoming of the protocol of Ref.~\cite{Lamport1982} in its original form is that it becomes exponentially data-intensive if a large number of cheating or unoperational nodes are present. 
Therefore further research on developing an efficient consensus protocol is required. 
We are optimistic that this issue can be resolved. 
Indeed, classical blockchain networks do routinely face the same challenge and have learned to deal with it efficiently~\cite{Croman2016}.

\section*{Appendix D. QKD network}

The basis for our experimental work is our recently developed modular QKD device~\cite{Sokolov2017,Kiktenko2016,Kiktenko2017,KiktenkoTrushechkin2016,Pozhar2017} 
driven by a National Instruments NI PCIe-7811R card. 
This setup uses a semiconductor laser LDI-DFB2.5G controlled by an FPGA board Spartan-6 to generate optical pulses at the standard telecommunication wavelength 1.55 $\mu$m 
and a 10 MHz repetition rate. 
We have used ID230 single-photon detectors from ID Quantique. 

The QKD network contains two links with different physical implementations, realized in an urban environment in Moscow. 
The parameters of both links are listed in the table~\ref{tab:2}. 

\begin{table}[H]
	\begin{center}
	\begin{tabular}{||c | c | c||} 
		\hline
		 & \bf{First link} & \bf{Second link}\\ [0.5ex] 
		\hline
		Encoding  & polarization & phase \\ 

		Length & 30 km & 15 km \\ 

		Loss & 13 dB & 7 dB \\

		QBER & $5.5\%$ & $3.5\%$\\

		Key rate & 0.02 Kbit/s & 0.1 Kbit/s \\

		\hline
	\end{tabular}
	\end{center}
	\caption{Main parameters of the links in the employed QKD network. QBER: quantum bit error rate}
	\label{tab:2}
\end{table}

\end{document}